\documentclass[manuscript]{emulateapj}
\usepackage{natbib,graphicx,amssymb,epsfig}

\slugcomment{ApJ Letters, in press}

\shorttitle{Ram pressure stripping in Pegasus}
\shortauthors{McConnachie et al.}

\begin{document}

\title{Ram Pressure Stripping of an isolated Local Group dwarf galaxy: evidence for an intra-group medium}
\author{Alan W. McConnachie$^1$, Kim A. Venn$^1$, Mike J. Irwin$^2$, Lisa M. Young$^3$, Jonathan J. Geehan$^1$}
\affil{$^1$Department of Physics and Astronomy, University of Victoria, Victoria, B.C., V8P 1A1, Canada}
\affil{$^2$Institute of Astronomy, Madingley  Road,  Cambridge, CB3  0HA,  U.K.}
\affil{$^3$Physics Department, New Mexico Tech., 801 Leroy Place, Socorro, NM 87801}

\begin{abstract}
We compare the stellar structure of the isolated, Local Group dwarf
galaxy Pegasus (DDO\,216) with low resolution HI maps from
\cite{young2003}. Our comparison reveals that Pegasus displays the
characteristic morphology of ram pressure stripping; in particular,
the HI has a ``cometary'' appearance which is not reflected in the
regular, elliptical distribution of the stars. This is the first time
this phenomenon has been observed in an isolated Local Group
galaxy. The density of the medium required to ram pressure strip
Pegasus is at least $10^{-5} - 10^{-6}$\,cm$^{-3}$. We conclude that
this is strong evidence for an inter-galactic medium associated with
the Local Group.

\end{abstract}


\section{Introduction}

\cite{einasto1974} first highlighted that dwarf satellites of large
galaxies tend to be gas deficient compared to isolated dwarfs. The
former generally have little or no ongoing star formation and the
stars are pressure supported (dwarf spheroidal, dSph). The latter
generally have ongoing star formation and the gas dynamics show that
rotational support is important (dwarf irregular,
dIrr). ``Transition'' dwarfs are gas-rich and, unlike dIrr galaxies,
have little or no detectable HII regions, although they usually show
indications of recent star formation.

The processes by which dwarf galaxies loose their gas are not fully
understood. Internal feedback, particularly winds from supernovae, are
likely important (\citealt{dekel1986}) and the existence of the
position-morphology relation clearly indicates that environmental
influences are significant. \cite{mayer2006} show that it is possible
for dwarf galaxies to be ram pressure stripped of some of their
gaseous component in a hot halo of the Milky Way or M31. This idea was
originally proposed by \cite{lin1983}, who calculated the density of
the medium required to be of order $10^{-6}$\,cm$^{-3}$. There
have been no direct detections of such a medium, although recently
\cite{nicastro2002,nicastro2003} and \cite{sembach2003} have detected
OVI absorption which they attribute to hot gas associated with either
a Milky Way corona or a Local Group medium.

In this {\it Letter}, we compare the stellar and gaseous structure of
the isolated, transition-type, dwarf galaxy Pegasus (DDO216). We show
that it displays the characteristic signature of ram pressure
stripping and conclude that this is strong evidence for hot gas
associated with the Local Group.  Table~1 summarises some of the
observed properties of Pegasus. We adopt the distance estimate by
\cite{mcconnachie2005a}, $D \simeq 919$\,kpc, derived from the same
photometry used in this Letter.

\begin{table}[htdp]
\begin{center}
\caption{Summary of observed parameters for the Pegasus (DDO216) dwarf galaxy}
\begin{tabular}{lll}
Parameter & Value & Reference \\
\hline
$\alpha$ (J2000)  &  23h 28m 36.2s& --- \\
$\delta$ (J2000) & +14$^\circ$ 44$^\prime$ 35$^{\prime\prime}$& --- \\
$(l, b)$ & $(94.8^\circ, -43.6^\circ)$ & --- \\
$M_V~{\it(L_V)}$ & -12.9~$(1.24 \times 10^7\,L_\odot)$ & \cite{mateo1998a} \\
$M_{HI}$ & $4.06 \times 10^6\,M_\odot$ & \cite{young2003}\\
$v_\odot$ & -183\,km\,s$^{-1}$ & \cite{young2003} \\
$v_r/\sigma$ & $1.7$ & \cite{mateo1998a} \\
Distance & 24.82 $\pm$ 0.07 (919\,kpc) & \cite{mcconnachie2005a} \\
         & 24.4  $\pm$ 0.2  & \cite{gallagher1998} \\
         & 24.9  $\pm$ 0.1  & \cite{aparicio1994} \\
\hline
\end{tabular}
\end{center}
\label{distances}
\end{table}
  
\section{Data}

On the night of 8 August 2003, we obtained Johnson V ($V^\prime$) and
Gunn i ($i^\prime$) photometry of Pegasus with the Wide Field Camera
(WFC) on the 2.5\,meter Isaac Newton Telescope (INT), a mosaic of four
CCDs with a total field of view of $27 \times 34$\,arcmins. This is
large enough that the entire dwarf galaxy is contained within a single
pointing. The seeing was typically $\lesssim 1.2"$ and exposure times
of 1000 seconds in each filter allowed us to reach $i^\prime \sim
23.5$\,mags and $V^\prime \sim 24.5$ with a signal-to-noise $\simeq
5$. These data were previously presented in \cite{mcconnachie2005a},
to which we refer the reader for more details. In the central regions
of Pegasus, crowding is severe and the photometry is very incomplete;
however, this makes no difference to any of the results in this {\it
Letter}.

The top-left panel of Figure~1 shows the reduced $V^\prime$ image of
Pegasus taken with the INT\,WFC. Also shown are the fields of view of
previous studies of Pegasus. In particular, the WIYN and HST WFPC2
fields analysed in \cite{gallagher1998} are shown in green as the
largest rectangular field and the small WFPC2 footprint; the NOT field
analysed by \cite{aparicio1994} is shown in blue as the smallest
rectangular field, and the field studied by \cite{hoessel1982} using
the 1.5m Palomar telescope is shown in red as the medium sized
rectangle.  As we show below, the extent of Pegasus is significantly
larger than has previously been recognised.

Low resolution HI data for Pegasus was presented in \cite{young2003}
and was obtained using the Very Large Array (VLA) with 2.2 hours in
the D array on 13 March 1995 and 12.7 hours in the C array on 24
January 1999. We refer the reader to \cite{young2003} for details of
the data reduction procedures.

\section{Analysis and Discussion}

\subsection{Comparison of stellar and HI contours}

The top-right panel of Figure~1 shows the tangent-plane projection of
the spatial distribution of objects identified as stellar from our
INT~WFC observations of Pegasus. The dotted lines in this panel (and
the remaining panels of Figure~1) correspond to the approximate edges
of each CCD of the INT~WFC. Only objects which lie within $1-\sigma$
of the stellar locus in both the $V^\prime$ and $i^\prime-$band
observations are shown. The hole at the center of the main body of
Pegasus is due to severe crowding which causes incompleteness. The
bottom-left panel of Figure~1 shows a contour map of the density
distribution of stars. The first contour is $2-\sigma$ above the
background, and the contours correspond to $2.2, 5.0, 8.6, 13.2, 19,0,
26.3, 35.7, 47.5$ and $62.5$\,stars\,arcmin$^{-2}$. The contour map
was made in the standard way and follows exactly the methodology
described in \cite{mcconnachie2006b}. This panel shows that Pegasus is
significantly more extended than suggested by the image in the first
panel.

The bottom-right panel of Figure~1 shows the stellar density
distribution as a grey-scale with square-root scaling. The red
contours are the low-resolution HI distribution from
\cite{young2003}. The contours correspond to column densities of
$0.1, 0.2, 0.4, 0.8, 1.6, 3.2, 6.4$ and $12.8 \times
10^{20}$\,cm$^{-2}$. Whereas the stars are distributed in a
regular ellipse (typical of a flattened spheroid or an inclined disk)
the HI has a ``cometary'' appearance; the contours in the south-east
are more closely packed and do not extend as far as in the north-west.

\subsection{Is Pegasus being ram pressure stripped?}

\begin{figure*}
  \begin{center}
    \includegraphics[angle=270, width=8.9cm]{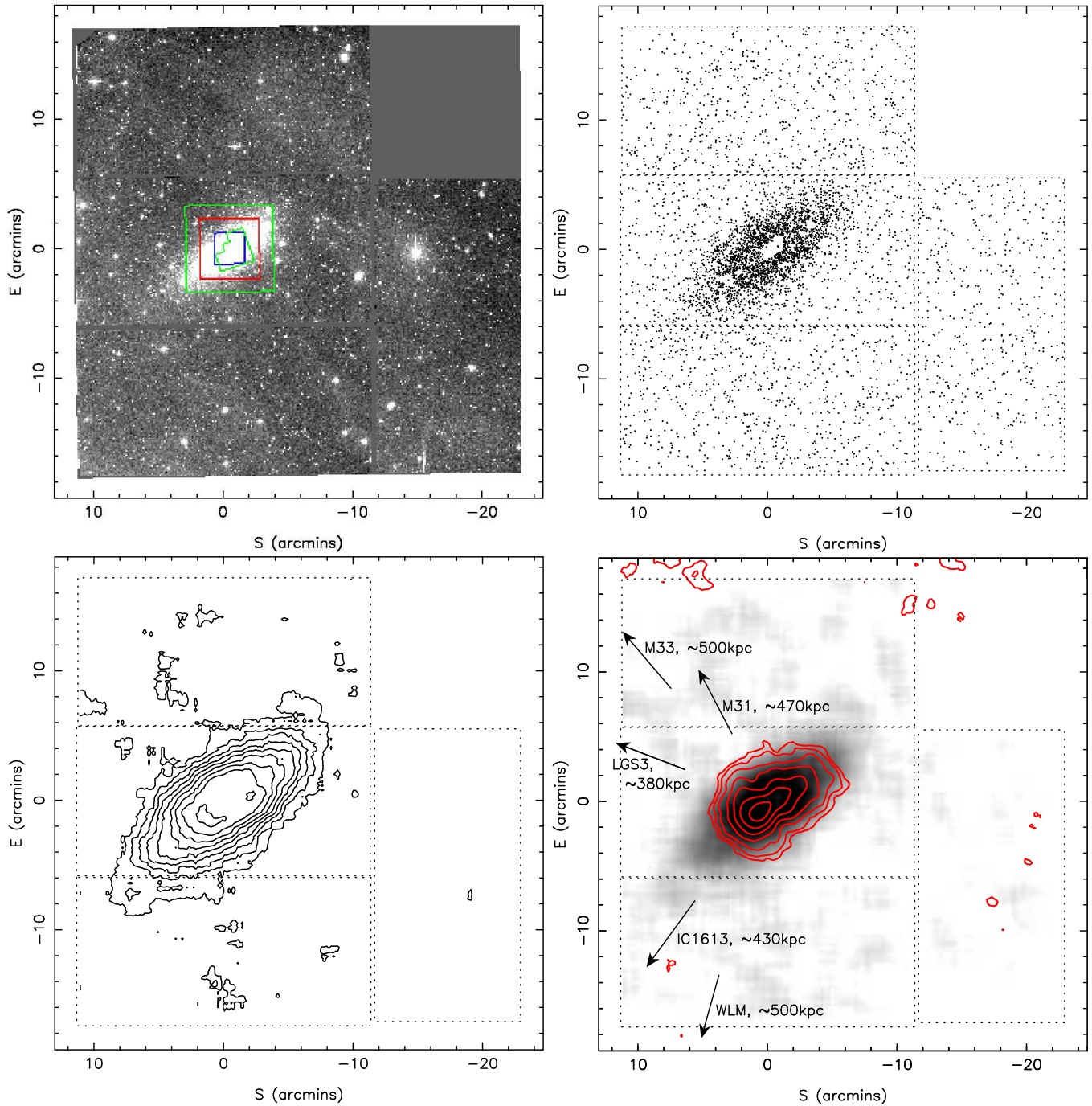}
    \includegraphics[angle=270, width=8.9cm]{f1b.ps}
    \includegraphics[angle=270, width=8.9cm]{f1c.ps}
    \includegraphics[angle=270, width=8.9cm]{f1d.ps}
    \caption{Projections in the tangent plane $\left(\xi, \eta\right)$
    of the structure of the Pegasus dwarf galaxy (DDO216) with the
    orientation of the field indicated. The dotted lines trace the
    approximate edges of the four CCDs of the INT~WFC. Top-left panel:
    The reduced $V^\prime-$band image of Pegasus taken with the
    INT~WFC. Also marked are the positions of fields analysed in
    previous studies of this galaxy; the fields analysed by
    \cite{gallagher1998} are shown in green as the largest rectangular
    field and the small WFPC2 footprint, the field analysed by
    \cite{aparicio1994} is shown in blue as the smallest rectangular
    field, and the field studied by \cite{hoessel1982} is shown in red
    as the medium sized rectangle. Top-right panel: The distribution
    of all sources confidently identified as stellar in both the
    $V^\prime$ and $i^\prime-$bands. The hole at the center of Pegasus
    is due to severe crowding causing the photometry to become
    seriously incomplete. Bottom-left panel: The stellar density
    distribution of Pegasus shown as a contour map. The first contour
    is $2-\sigma$ above the background and the 9 contours correspond
    to $2.2, 5.0, 8.6, 13.2, 19,0, 26.3, 35.7, 47.5$ and
    $62.5$\,stars\,arcmin$^{-2}$. Bottom-right panel: the stellar
    density distribution is shown as a grey-scale with square-root
    scaling. The red contours show the low resultion HI distribution
    from \cite{young2003}. Contours correspond to $0.1, 0.2, 0.4, 0.8,
    1.6, 3.2, 6.4$ and $12.8 \times 10^{20}$\,cm$^{-2}$. Also
    shown are the projected directions to all galaxies within $\sim
    500$\,kpc of Pegasus which have a significant gaseous
    content. Pegasus displays the characteristic morphology of ram
    pressure stripping.}
    \label{fig}
  \end{center}
\end{figure*}

The shape of the low resolution HI contours in Pegasus - the smooth,
compressed contours in the south-east and the ``tail'' to the
north-west - is very similar to the simulated morphology of gas
undergoing ram pressure striping (e.g.,
\citealt{stevens1999,mori2000,marcolini2003,roediger2005,mayer2006}).
Observationally, the M81 group dwarf galaxy Holmberg~II is observed to
have a similar morphology (\citealt{bureau2002}), interpreted as
evidence of an intra-group medium. In clusters of galaxies, ram
pressure stripping of galaxies by an intra-cluster medium is used to
explain various observations, including the deficit of HI in cluster
spiral galaxies compared to field spirals (e.g.,
\citealt{giovanelli1985}). Indeed, several individual galaxies in the
Virgo Cluster have been shown to display gaseous morphologies
indicative of ram-pressure stripping
(\citealt{vollmer2000,vollmer2004,vollmer2005}).

What else could explain the peculiar appearance of Pegasus? While
tidal stripping by large galaxies can affect the structure of dwarf
galaxies (e.g., \citealt{penarrubia2007b}), the closest large galaxy
to Pegasus is M31 at $\sim 470$\,kpc (all the distance estimates in
Table~1 place Pegasus at $> 400\,$kpc from M31).  Even if we assume
Pegasus is a weakly-bound satellite of M31, tidal effects at this
distance are minimal. If Pegasus was disrupted at pericenter, it is
unlikely that the gas would still show signs of current
disturbance. Further, tidal stripping tends to produce symmetrical
distortions and both gas and stars should be affected. However, these
are inconsistent with the structure of Pegasus that we observe.

Could the appearance of Pegasus be due to internal effects rather than
external influences? Enhanced star formation in the south-east of
Pegasus could produce winds which remove gas from this
region. However, if this is the case then the densely packed contours
in the south-east should have a more concave, rather than convex,
shape. For example, \cite{young2007} discuss a gas cloud associated
with the Phoenix dwarf galaxy and conclude that it was blown out by
supernovae winds based in part on the concave shape of its contours.

An alternative explanation for the HI morphology of Pegasus is that it
consists of multiple HI clouds, the sum total of which has a cometary
appearance. Figure~6 of \cite{young2003} is a position-velocity
diagram of Pegasus along its major axis. It shows a gradient in
velocity and two main concentrations of HI which \cite{young2003}
interpret as two distinct HI clouds. The strength of the secondary
feature ($v \sim -200$\,km\,s$^{-1}$) is weaker than the main feature
($v \sim -180$\,km\,s$^{-1}$) and they join at relatively high column
density (between the $8 - 16 - \sigma$ contour levels). An alternative
explanation of the data is that the overall velocity gradient is a
result of ram-pressure stripping. The velocity difference between the
two features may be due to stripped gas leaving a ``hole'' in the
distribution, making the secondary feature appear at a higher density
than its immediate surroundings (we do not necessarily expect that the
column density should smoothly vary over the entire cloud).

Henceforth, we adopt the hypothesis that Pegasus is being ram pressure
stripped. Following \cite{gunn1972}, material will be ram pressure
stripped from a galaxy if the density of the surrounding medium,
$n_{IGM} \gtrsim
\left(2\,\pi\,G\,\Sigma_{T}\,\Sigma_{HI}\right)/\left(\mu v^2\right)$.
$\Sigma_{T}$ is the total surface density (stars plus gas),
$\Sigma_{HI}$ is the column density of HI and $v$ is the relative
velocity of the galaxy to the medium. Thus,

\begin{eqnarray}
n_{IGM} &\simeq& 3.7 \times 10^{-6}\,\rm{cm}^{-3}
\left(\frac{\rm{100\,km\,s^{-1}}}{v}\right)^2 \nonumber\\
&&\left(\frac{\Sigma_{HI}}{0.1 \times 10^{20}\,\rm{cm}^{-2}}\right)^2~,
\end{eqnarray}

\noindent where we take the mean particle mass $\mu = 0.75\,m_p$ for
fully ionized media.  We approximate the Local Group space velocity of
Pegasus as $v \sim \sqrt{3}\,\sigma_{LG} \sim 100$\,km\,s$^{-1}$ where
$\sigma_{LG} \sim 60$\,km\,s$^{-1}$ is the Local Group line-of-sight
velocity dispersion (\citealt{sandage1986}).  HI at a column density
much lower than $\Sigma_{HI} \sim 0.1 \times 10^{20}$\,cm$^{-2}$ has
been stripped from Pegasus, implying that this is a reasonable lower
limit for use in this calculation. We adopt $\Sigma_T = \Sigma_{HI}
\left(1 + M_\star/M_{HI}\right)$, where $M_\star \sim 1.24 \times
10^7\,M_\odot$ is the stellar mass of Pegasus (Table~1). This seems
reasonable; the surface brightness of Pegasus is
$25$\,mags\,arcsec$^{-2}$ at a radius of $r = 1.5^\prime$ on the minor
axis (\citealt{nilson1973,devaucouleurs1991}), corresponding to a
stellar surface density of $\Sigma_\star \sim 4 \times
10^{20}$\,cm$^{-2}$. This is approximately equivalent to the
stellar-to-gas mass ratio multiplied by the gas surface density
($M_\star/M_{HI} \times \Sigma_{HI}$) at $r = 1.5^\prime$.

These values yield $n_{IGM} \sim 3.7 \times
10^{-6}$\,cm$^{-3}$. However, given the uncertainties involved, it is
entirely plausible that the value of $n_{IGM}$ could be at least an
order of magnitude larger than in Equation~1.

\subsection{Consequences}

What is the source of the material that is stripping Pegasus? The
bottom right panel of Figure~1 shows the distances of Pegasus to its
nearest gas-rich neighbours. The dwarf neighbours are unlikely to be
the source of the stripping medium; not only is the required mass of
gas unrealistically large (an ejected spherical shell $\sim 1$\,kpc
thick with a radius of $\sim 300$\,kpc would have a mass $> 3 \times
10^6\,M_\odot$ at a density of $n_{IGM}$) but the energy required is
too large for a dwarf galaxy to reasonably provide.

Alternatively, the gas could be associated with M31. From observations
of the Magellanic stream, \cite{murali2000} estimate that the density
of the Milky Way halo at the stream must be $\lesssim
10^{-5}$cm$^{-3}$, although \cite{stanimirovic2002} estimate $\sim
10^{-4}$ cm$^{-3}$. If the gas density in the halo of M31 is
similar, then not only would M31 need to have a very extended corona,
but its density would need to decrease very slowly with
radius. Indeed, if the Milky Way has a similarly extended corona, then
the two will overlap and the result may be observationally
indistinguishable from a Local Group medium.

The isolation of Pegasus raises the strong possibility that the
stripping medium is associated with the Local Group, rather than
individual galaxies within the group. Clusters of galaxies have such
media, and observations of Holmberg~II imply the presence of an
intra-group medium in the M81 group (\citealt{bureau2002}). The
density of the intra-group medium implied in Equation~1 is of the same
order as the density of the medium responsible for local OVI
absorption detected by \cite{nicastro2002,nicastro2003} and
\cite{sembach2003}, which they suggest is associated with either a
Milky Way corona or a Local Group medium. Our result favors the
latter interpretation. Theoretically, $\sim 30\,\%$ of baryons in the
Local Volume are expected to be in a warm/hot phase ($T \sim 10^5 -
10^6\,$K; \citealt{kravtsov2002}); this is likely concentrated around
galaxies and galaxy groups as an intra-group medium.

If the stripping medium pervades the Local Group, why do more dwarf
galaxies not show evidence of ram pressure stripping? \cite{lin1983}
suggest that all the dSphs have been stripped in this fashion,
(although \cite{mayer2006} show that ram-pressure stripping by itself
is insufficient to remove all the gas from a dIrr). It is possible
that the Local Group medium will be clumpy and perhaps Pegasus is
passing through a region of higher density compared to other
dIrrs. Alternatively, Pegasus could be falling into and interacting
with the Local Group for the first time, as has recently been
speculated for two dSph galaxies at large radii from M31 (And~XII,
\citealt{chapman2007}; AndXIV, \citealt{majewski2007}). However, the
reason why only Pegasus currently shows signs of ram-pressure
stripping is unlikely to be known until such time as the masses and
orbits of the dIrrs have been determined. Given the distances of these
galaxies, this will be some time yet.

\section{Summary}

We show that the isolated, transition-type, Local Group dwarf galaxy,
Pegasus (DDO216) is undergoing ram pressure stripping. We calculate
that the density of the medium required to strip Pegasus is at least $n_{IGM}
\sim 10^{-5} - 10^{-6}$\,cm$^3$, of the same order as the medium
recently identified by \cite{nicastro2002,nicastro2003} and
\cite{sembach2003} through OVI absorption. Given the large distance of
Pegasus from either the Milky Way or M31, we conclude that Pegasus
presents strong evidence for the existence of a Local Group
inter-galactic medium.

\acknowledgements{We thank Mary Putman, Stephanie C{\^o}t{\'e}, Evan
Skillman, Andi Mahdavi, Arif Babul, and Chris Bildfell for valuable
conversations. AWM is supported by a Research Fellowship from the
Royal Commission for the Exhibition of 1851, and thanks J.~Navarro and
S.~Ellison for additional financial assistance. KAV and JG thank NSERC
for support through a Discovery grant.}

\end{document}